**A Guiding Framework for K-12 Teachers in Creating AI-powered Learning Technologies through Vibe Coding**

Yukyeong Song[a]
Assistant Professor
E-mail: ysong51@utk.edu

Seoyeon Choi[b]
Ph.D Student
Email: smdus@korea.ac.kr

Jinhee Kim[c]
Assistant Professor
Email: jhkim@odu.edu

Yeongje Kim[b]
Master's Student
E-mail: xkxh2004@korea.ac.kr

Lauren Weisberg[d]
Assistant Professor
E-mail: lauren.weisberg@uta.edu

Jewoong Moon[e]
Assistant Professor
E-mail: jmoon19@ua.edu

***Corresponding author**: Yukyeong Song is an Assistant Professor in the Theory and Practice in Teacher Education at the University of Tennessee, Knoxville. Her research focuses on artificial intelligence education, designing AI-powered learning technologies, and learning analytics.

[a] Department of Theory and Practice in Teacher Education, University of Tennessee, Knoxville, TN 37916, USA

[b] Department of Education, College of Education, Korea University, Seoul 02841, South Korea.
[c] Department of STEM Education and Professional Studies, Old Dominion University, Norfolk, VA, USA

[d] Department of Teacher and Administrator Preparation, University of Texas, Arlington, TX, 76019, USA

[e] Department of Educational Leadership, Policy, and Technology Studies, The University of Alabama, Tuscaloosa, AL 35401, USA

## Abstract

Large language models generate code from natural language prompts, enabling “vibe coding,” which allows non-programmers to develop computational solutions. Vibe coding for teachers amplifies the value of *teachers-as-designers*, improving technology integration while fostering AI literacy. However, structured guidance on supporting this process is lacking. We propose *GAIDE (A Guiding Framework for AI-Integrated Design for Educators)*, a framework that supports K-12 teachers in creating AI-powered learning technologies through vibe coding. The initial framework, built on Design Thinking and INTERACT, was validated through a CORDTRA interaction analysis of three teachers and four faculty mentors in an eight-week workshop to derive the final framework. Additionally, the qualitative analysis of pre- and post-interviews found an enhancement of teachers’ AI literacy. Findings highlight the potential of learning-by-creating for professional development.

**Keywords:** teachers-as-designers, vibe coding, AI-powered learning technology creation, AI literacy, professional development

# I. Introduction

Recent advances in generative artificial intelligence (GenAI) have begun to transform how software is created (Alenezi & Akour, 2025). Large language models (LLMs) are increasingly capable of generating functional code from natural language prompts, enabling a practice that has recently been referred to as *"vibe coding,"* a development approach where users describe desired functionality in natural language and AI generates, revises, and executes the underlying code (Karpathy, 2025). In this paradigm, users guide development through high-level intent rather than through traditional programming syntax, lowering the barrier to entry for software development (Chow & Ng, 2025). As a result, individuals without formal programming training, including domain experts, educators, and practitioners, can independently develop tailored computational solutions with contextual expertise.

In education, this shift has important implications. Many educational tools developed mainly by developers or researchers are criticized for being poorly aligned with classroom realities (Luckin et al., 2019). Recognizing this problem, co-design and participatory approaches have sought to involve teachers as co-designers of learning technologies to ensure pedagogical relevance and usability (Roschelle et al., 2006). Research has shown that engaging teachers in design processes can strengthen their pedagogical reasoning about technology and increase the likelihood that innovations are successfully adopted in classroom contexts (Kali et al., 2015; Penuel et al., 2007).

Another advantage of engaging teachers in technology development is that it can enhance teachers' technological knowledge and competency (Kali et al., 2015). The growing integration of AI into education has generated an emphasis on developing teachers' AI literacy (Ding et al., 2024). AI literacy refers to the knowledge, skills, and critical understanding needed to interpret, evaluate, and effectively use AI systems in professional and societal contexts (Long & Magerko, 2020; Ng et al., 2021). For educators, AI literacy is increasingly important for responsible and pedagogically meaningful use of AI in teaching and learning (Pei et al., 2025). One promising pathway for fostering AI literacy is a project-based development process (Kali et al., 2015; Papert, 1980). From the perspectives of *learning by design* (Koehler & Mishra, 2005), enabling teachers to build AI-powered learning technologies may serve as a mechanism for creating contextualized classroom tools and a professional learning experience that deepens their understanding of AI.

Despite these opportunities, teachers who wish to create AI-powered technologies face multiple challenges. While technological challenges have largely been mitigated by AI-assisted coding, teachers could face limited technical capacity and professional development opportunities (Celik et al., 2022). In addition, developing educational software requires not only content and pedagogical knowledge but also the ability to navigate design decisions related to data, interaction design, and ethical considerations (Domagk et al., 2010). Many teachers have limited experience in these areas, and existing professional development programs rarely provide structured guidance in building AI-enabled learning technologies. While mentorship from university researchers or experienced educators can provide support (e.g., Koehler & Mishra, 2005; Song et al., 2026), there remains little research on how to effectively guide teachers in creating AI-powered teaching and learning tools while they retain a high level of agency.

In this paper, we address these gaps by proposing a guiding framework to support K-12 teachers in creating AI-powered learning technologies through AI-assisted vibe coding. We named our new framework GAIDE (A Guiding framework for AI-integrated Design for Educators). With the initial framework drawn from existing Design Thinking and INTERACT frameworks, we conducted an eight-week design workshop with three teachers and four university faculty mentors to validate the practical usability of our framework. In addition, we investigate teachers' design activities and mentors' support activities demonstrated in each stage of the design process through CORDTRA analysis. Finally, we analyzed the pre- and post-interviews to explore how teachers' AI literacy evolved throughout the workshop.

This study addresses the following research questions (RQs):

**RQ1.** How is the teachers' AI-powered learning technology creation process demonstrated in the workshop?

**RQ2.** What activities do teachers and faculty mentors demonstrate during the guided AI learning technology creation process?

**RQ3.** How does K-12 teachers' AI literacy evolve throughout the AI-powered learning technology creation process?

# II. Literature Review

## 2.1. AI-assisted Vibe Coding in Education

*Vibe coding* refers to a flow-oriented programming through conversational prompting to AI to generate, refine, and debug code (Karpathy, 2025). Vibe coding emphasizes humans in the loop as the central actors in executing the code, comparable to agentic coding, which allows autonomous agents to operate with minimal human intervention (Kusper & Szabó, 2025). Vibe coding has made programming more accessible, allowing people with limited programming knowledge and experience to build functional apps that meet their needs. In a recent AI-assisted coding competition, winners included subject-matter experts in different domains, rather than professional programmers, such as physicians, lawyers, and musicians, who created the tailored solutions grounded in their domain knowledge (Minervee, 2026).

Recent studies have begun to explore its potential in education. Chow and Ng (2025) adopted vibe coding in medical education and reported two clinical applications created by a medical educator with no formal programming training. In higher education, Kusper and Szabo (2025) examined the use of vibe coding in a college engineering course and found that "teaming" with AI during programming tasks significantly improved students' self-reported productivity, although it did not increase their confidence in the correctness of their code. In K-12 education, Woo and colleagues (2025) presented contrasting cases of secondary school students developing English-as-a-foreign-language learning tools through vibe coding in a four-hour workshop. While these examples illustrate the emerging exploration of vibe coding in education, little research has examined how teachers can create classroom solutions through vibe coding.

### 2.2. Developing Teachers' AI Literacy through Technology Creation

As AI becomes increasingly embedded in education, developing teachers' AI literacy has emerged as a priority (Pei et al., 2025). AI literacy often encompasses the four major components, including the abilities 1) to know and understand AI, 2) to use and apply AI, 3) to evaluate and create AI, and 4) AI ethics (Ng et al., 2021). For educators, this includes not only technical awareness but also the ability to make informed pedagogical decisions, critically assess AI outputs, and understand ethical and societal implications of AI use in classrooms (Ding et al., 2024).

One promising approach to fostering such understanding is grounded in Constructionist learning theory, which posits that learners develop deeper knowledge through the process of actively creating meaningful artifacts (Papert, 1980). This perspective has been operationalized through *learning-by-design* (Kalantzis & Cope, 2005; Koehler & Mishra, 2005) and *learning-by-making* (Hsu et al., 2017) approaches, where designing and building technologies serve as a form of professional learning. Prior work has shown that engaging teachers in the design of learning technologies can enhance their technological knowledge and literacy (Kali et al., 2015; Koehler & Mishra, 2005). Extending this line of research, enabling teachers to create AI-powered tools holds potential to provide a powerful pathway for developing AI literacy by situating learning within authentic, goal-directed design activities. Despite this potential, there are limited efforts to support teachers in creating AI-powered learning technologies and to examine the impact of such experience on developing teachers' AI literacy.

### 2.3. Guiding Frameworks for Learning Technology Design

Research has proposed guiding frameworks to support the systematic design of learning technologies and tools (e.g., Richey & Klein, 2013), aiming to bridge theory and practice by providing structured approaches that help designers translate pedagogical principles into effective technological solutions. Among many such frameworks that emphasize iterative, user-centered design processes alongside theoretically grounded considerations of how learners interact with and benefit from technology, Design Thinking has been widely adopted to guide solution design and development.

Design Thinking is a human-centered, iterative approach that supports the creation of innovative and practically useful solutions through five interconnected phases: empathizing with users, defining problems, ideating solutions, prototyping, and testing through user feedback (Razzouk & Shute, 2012). Its emphasis on understanding users' needs, considering social contexts, and supporting rapid prototyping has led to its increasing adoption in AI education research. For example, Song et al. (2023) engaged middle school students in Designing Thinking to create conversational AI applications that address real-world needs, foregrounding the social impact of AI as highlighted in the AI4K12 initiative (Touretzky et al., 2019). While Design

Thinking provides a generic procedural structure for guiding the development process (e.g., needs analysis, prototyping, and testing), it does not specify what makes the resulting AI interaction pedagogically effective.

Meanwhile, the INTERACT framework offers a perspective for designing interactive multimedia environments that support meaningful cognitive engagement (Domagk et al., 2010). The framework provides a conceptual bridge between instructional theory and the design decisions teachers must make when configuring AI-mediated learning, guiding teacher-led AI development by translating theoretical conditions into actionable design components (Ding et al., 2024). For example, the framework proposes the main components of interactivity design, including 1) learning environment, 2) behavioral activities, 3) cognitive and metacognitive activities, 4) emotions and motivation, 5) mental model, and 6) learner variables. In addition, the framework emphasizes the considerations in learning objectives, instructional events, interactive affordances, and support mechanisms. In this sense, INTERACT complements Design Thinking by providing guidance on designing pedagogically effective interactions. Therefore, we integrated Design Thinking and the INTERACT framework to build the initial theoretical model for our guiding framework (Figure 1).

**Figure 1**

*Initial process model*

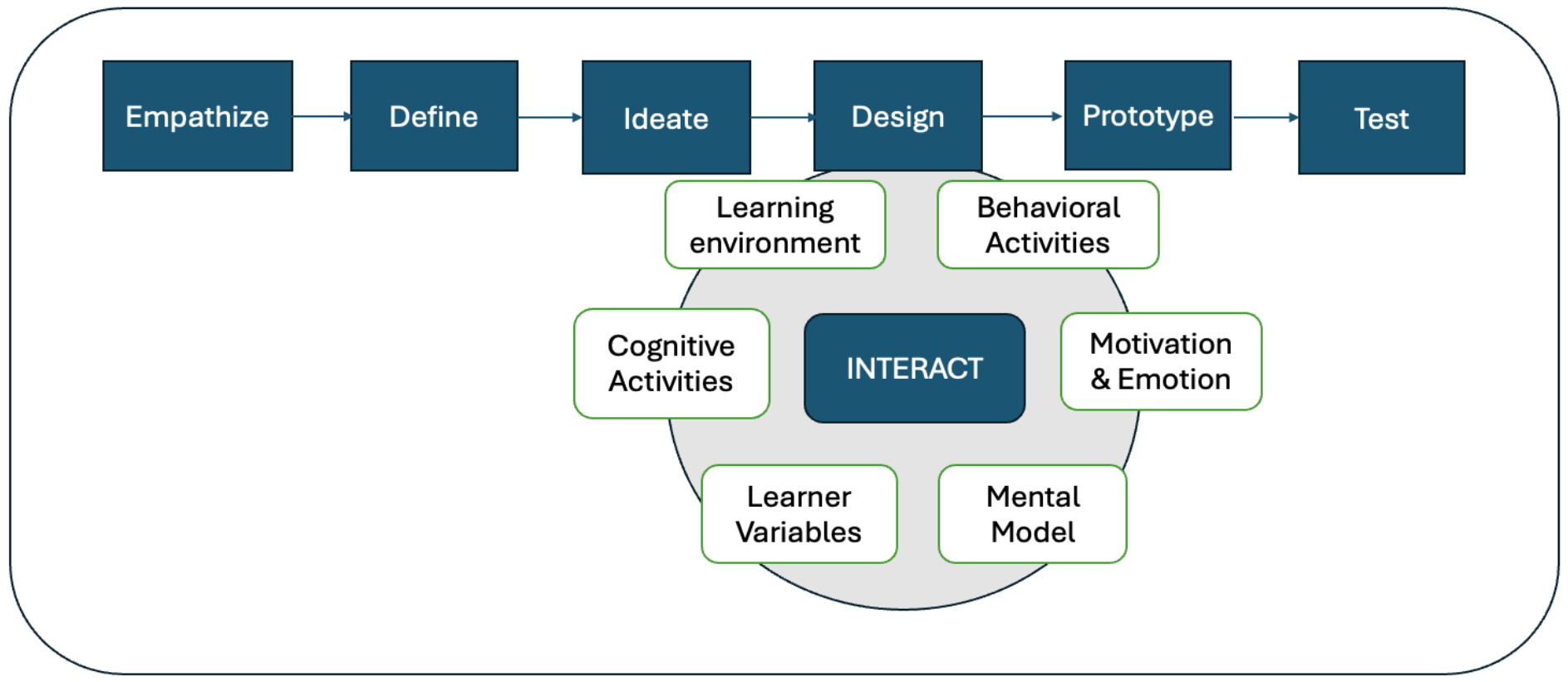

# III. Methods

## 3.1. K-12 Teacher Workshop: Empirical Validation of the Initial Framework

This study proposes a new guiding framework, GAIDE, supported by an empirical implementation. We conducted an eight-week-long, faculty-mentored synchronous online (Zoom) workshop designed to guide K-12 teachers in designing and developing AI-powered classroom solutions. This workshop offers teachers a situated learning opportunity by supporting them in competing in the U.S. Presidential AI Challenge. As part of the Executive Order 14277, the Presidential AI Challenge offers "a national competition where K-12 youth, educators, mentors, and community teams come together to solve real-world problems in their communities using AI-powered solutions" (The White House et al., 2025). Although the Challenge provides a high-stakes and authentic context for engaging with AI, it is primarily structured as a competition for independent teams and offers limited instructional scaffolding for educators with minimal prior experience in AI design and development. Our workshop was intentionally designed to address this gap by providing structured pedagogical and technical guidance for teacher-led AI innovation.

Over the eight weeks, the teacher team collaboratively created *Math Mysteries*, a storytelling-based, AI-powered, gamified platform designed to support middle school mathematics learning (Figure 2). Teachers used Google's AI Studio for prototyping through vibe coding. AI Studio was selected because it is free for educators, operates entirely in the browser without requiring installation, and supports multimodal AI generation, including text, images, video, audio, and code, through Gemini-family models.

**Figure 2**

*The Interface of Math Mysteries: Teacher-developed AI-powered Learning Technology*

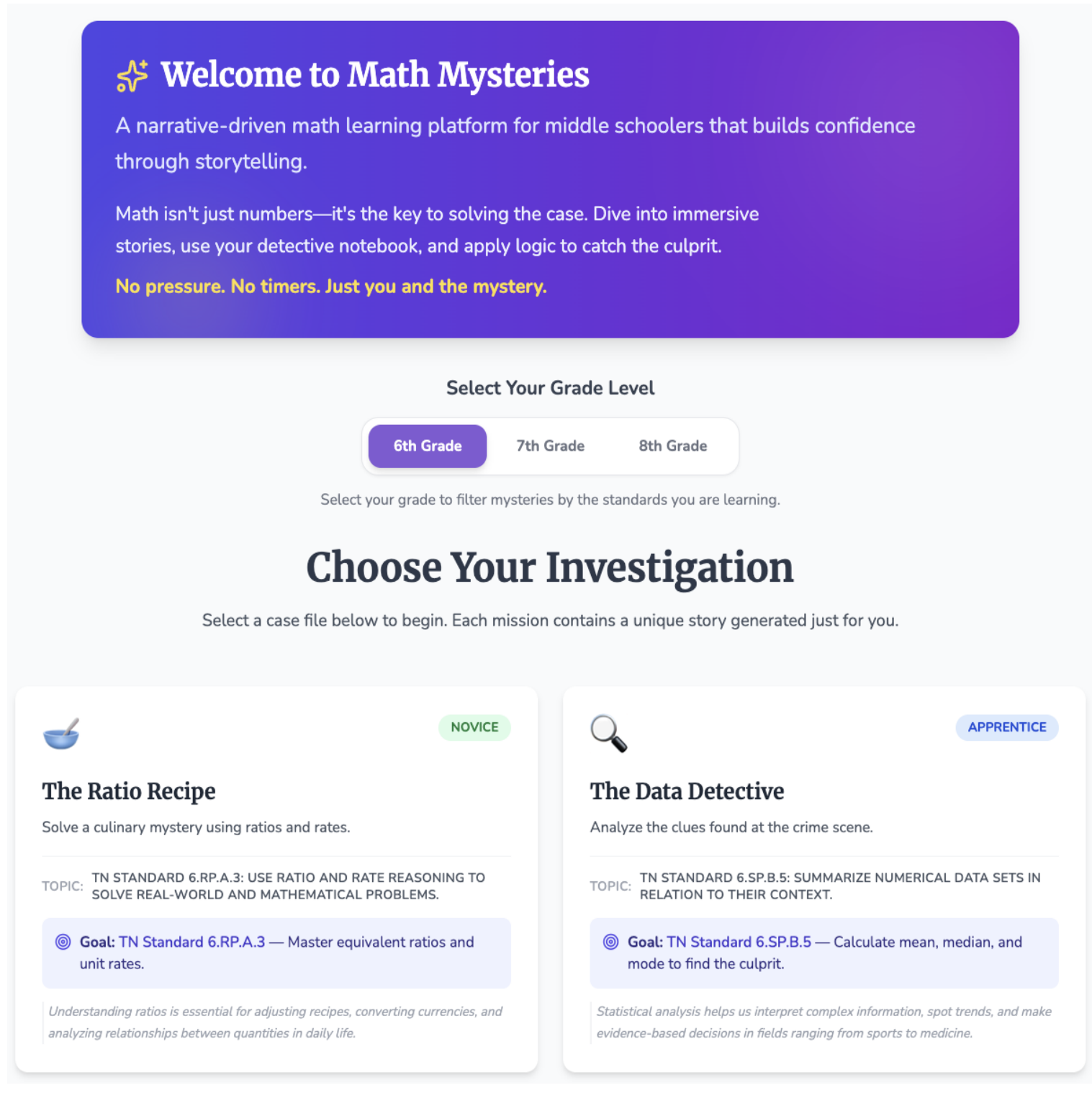

Session one focused on orientation, team building, an overview of the Challenge requirements, and the pre-interview. The second session began with a mentor-led mini-lecture introducing foundational concepts of AI. From sessions three through seven, teachers collaboratively ideated, designed, and developed AI-powered solutions guided by our initial framework (Fig. 1). Beginning in session five, portions of workshop time were allocated to Challenge preparation, with session eight dedicated entirely to this purpose. Due to time constraints, teachers were also asked to continue prototyping outside of the sessions. Post-interviews were conducted following session eight.

The participants included three K-12 teachers and four university faculty mentors (Table 1). The three teachers were from rural East Tennessee and participated as designer-practitioners voluntarily responding to the Presidential AI Challenge. Four faculty members from three research-intensive (R1) universities in the southern United States served as mentors for the teacher team. Both teachers and mentors were recruited through snowball sampling, with the first author initiating recruitment by contacting potential participants.

**Table 1**
*Participant Information*

| Participants | School-level | Subject or Major | Experience |
|---|---|---|---|
| **Teacher A** | Middle school (Librarian) | Vocabulary, Digital Citizenship | 12 years |
| **Teacher B** | Middle school (6th) | History, Computer science | 5 years |
| **Teacher C** | High school (9-10th) | English, ELA | 11 years |
| **Faculty A** | R1 Institute | Educational Technology | 1 year |
| **Faculty B** | R1 Institute | Learning, Design, and Technology | 5 years |
| **Faculty C** | R1 Institute | Computer Science | 22 years |
| **Faculty D** | R1 Institute | Educational Technology | 5 years |

### 3.2. Data Collection

The major collected data include 1) workshop video recordings and 2) pre- and post-interviews. The eight sessions of the weekly online workshop were audio-, video-recorded and transcribed using the in-platform AI transcription service on Zoom, which was reviewed by a human researcher and manually corrected as needed. Videos were recorded to capture any screenshares and provide contextual information. Pre- and post-focus group interviews allowed three teachers to take turns responding to the interview questions. The pre-interview included questions on the prior AI knowledge (e.g., "How would you explain what AI is to your students?"), prior experience (e.g., "What kinds of AI tools have you used in your teaching?"), perspectives on AI in education (e.g., "What do you see as the main opportunities and challenges of AI in education?"), and self-efficacy beliefs (e.g., "How confident do you feel about learning to create AI-powered learning technology?"). After all sessions, a post-interview was conducted to examine participants' reflections on their learning experiences (e.g., "How would you describe your overall experience?"), perceived growth in AI literacy (e.g., "How would you now explain what AI is to your students?"), and perspectives on integrating the developed AI tools into their classroom practice (e.g., "How do you envision using the tool you developed in the classroom?"). In addition, during the post-workshop interview, we shared our screen to show our initial model used for the workshop design (Fig. 1) and asked the teachers for their opinions on the visualization and terminology used in the model. Ethical approval for this study was obtained from the Institutional Review Board at the first author's institute.

### 3.3. Data Analysis

#### *3.3.1. Video analysis using CORDTRA diagram*

The recorded workshop sessions were analyzed to reveal the empirical process of AI-powered tool design and creation (RQ1) and participating teachers' and mentors' interpersonal and human-AI interactions (RQ2). To do that, we used CORDTRA (Chronologically Ordered Representation of Discourse and Tool Related Activity) diagrams (Hmelo-Silver et al., 2008), which visualize chronological interactions by presenting multiple codes of actions, such as utterances, tool use, and gestures, revealed from data over time in parallel within a single table (Hmelo-Silver et al., 2009).

CORDTRA diagrams represent time on the x-axis and data categories on the y-axis, with data patterns displayed within the diagram (Hmelo-Silver et al., 2008), replicating Choi et al. (2026)'s diagram-construction method. Accordingly, in this study, time was marked at one-minute intervals on the x-axis, the derived codes were listed on the y-axis, and data patterns were indicated by different colors. We only coded six sessions out of the total eight sessions, as sessions 7 and 8 were fully dedicated to preparing materials for the Presidential AI Challenge submissions (e.g., creating demo videos, writing applications). Sessions 4, 5, and 6 were also partially used for the material development, so we only coded the parts of the video recordings that are related to the design and development of the AI-powered learning technology.

#### *3.3.2. Coding scheme*

For the CORDTRA analysis, we developed a coding scheme to systematically analyze teachers' and mentors' utterances and activities (Appendix A). We employed a hybrid coding approach combining inductive and deductive methods (Fereday & Muir-Cochrane, 2006). The coding scheme for analyzing teachers' dialogic moves was derived from the literature focusing on cognitive processes in task-based collaborative lesson design (e.g., Yan and Goh, 2023) to capture key interactional patterns observed in collaborative discourse. In addition, codes related to design actions and tool use that emerged in teachers' interactions with AI (e.g., Prompt Writing, Demonstrating Prototype, Refining Ideas, Reviewing AI Outputs) were inductively derived from the data.

For the mentors' activities, the coding scheme was developed based on the previous literature on tutoring, scaffolding, and facilitating (e.g., Hmelo-Silver & Barrows, 2006; Song et al., 2025). In addition, to reflect the AI-based design context, we inductively added new codes, such as Technological Walkthrough and Troubleshooting Support. For these inductive codes, two researchers pilot-coded approximately 10-minute segments from eight video recordings and engaged in iterative review and discussion to refine and generate new codes.

To enhance the systematic organization and traceability of the coding process, we transcribed and organized the workshop videos using a Google Spreadsheet. The transcripts were structured at the level of individual turns, with separate columns distinguishing teachers and mentors. Based on these transcripts, two researchers assigned codes directly in Google Sheets while simultaneously coding using the CORDTRA diagram. When marking data patterns in the CORDTRA diagrams, two coders watched all the recorded videos together and discussed until they reached 100% agreement (Choi et al., 2026).

#### *3.3.3. Interview Analysis*

To investigate participating teachers' evolving AI literacy over time, we conducted a thematic analysis of transcribed pre- and post-interviews using Braun and Clarke's (2006) six-step framework. Two researchers independently reviewed the transcripts and generated initial codes reflecting four domains of AI literacy (Ng et al., 2021): *know and understand AI*, *use and apply AI*, *evaluate and create AI*, and *AI ethics*, noting shifts in participants' perceptions before and after the workshop. Axial coding grouped codes into categories, capturing key aspects of AI literacy development. Themes were iteratively refined by comparing coded data across participants and timepoints, highlighting how competencies within each domain evolved. Representative quotes and contextual details were documented to enhance credibility and transferability. To ensure trustworthiness, member checking was conducted with participants, and iterative discussions continued until consensus was reached among researchers, participants, and expert reviewers.

## IV. Results

### 4.1. Teachers' Technology Creation Process Demonstrated in the Workshop

Figure 3 illustrates the CORDTRA diagram of the teacher workshop. Each diagram represents one workshop session (Fig. 3-(a)−(c)), while the final diagram (Fig.3-(d)) integrates sessions 5, 6, and 7. The top row displays the color-coded "Steps in Developing." Below, teachers' and mentors' minute-by-minute activities are annotated using the coding scheme, with colors corresponding to the respective development steps.

In Figure 3-(a), the process was largely centered on the *Understand AI* phase, with the mentor introducing foundational AI concepts. In Figure 3-(b), the process extended *Empathize & Analyze*, where teachers examined the problem context. Then it moved to *Define Educational Problem* to clarify the issue, followed by the *Ideate* phase to explore possible solutions. In Figure 3-(c), the process exhibited a more iterative pattern. *Understand AI* reappeared in a more applied form, with hands-on guidance in Google AI Studio, followed by the *Design* phase, based on the INTERACT framework, where teachers discussed and made decisions on *Learning Objectives (LO)*, *Learner Differentiation (LD)*, and *Knowledge Mapping (KM)*. Teachers revisited *Empathize & Analyze*, *Ideate*, and *Define Educational Problem* as they refined their ideas and developed a *Prototype*. Between sessions, teachers developed the prototypes individually and collaboratively outside of the workshop as assignments. In Figure 3-(d), teachers tested their prototypes and received mentor feedback. They briefly revisited *Define Educational Problem* and *Empathize & Analyze* to refine their ideas, then continued with *Interactive Design*, incorporating *Behavioral Activities (BA)*, *Learner Differentiation (LD)*, and *Thinking Process (TP)*. This cycle of prototyping and feedback continued, eventually leading to the *Reflect on Ethical Application* phase.

**Figure 3**

*CORDTRA Diagram of Teachers' AI Learning Technology Creation Process*

*(a) Session 2*

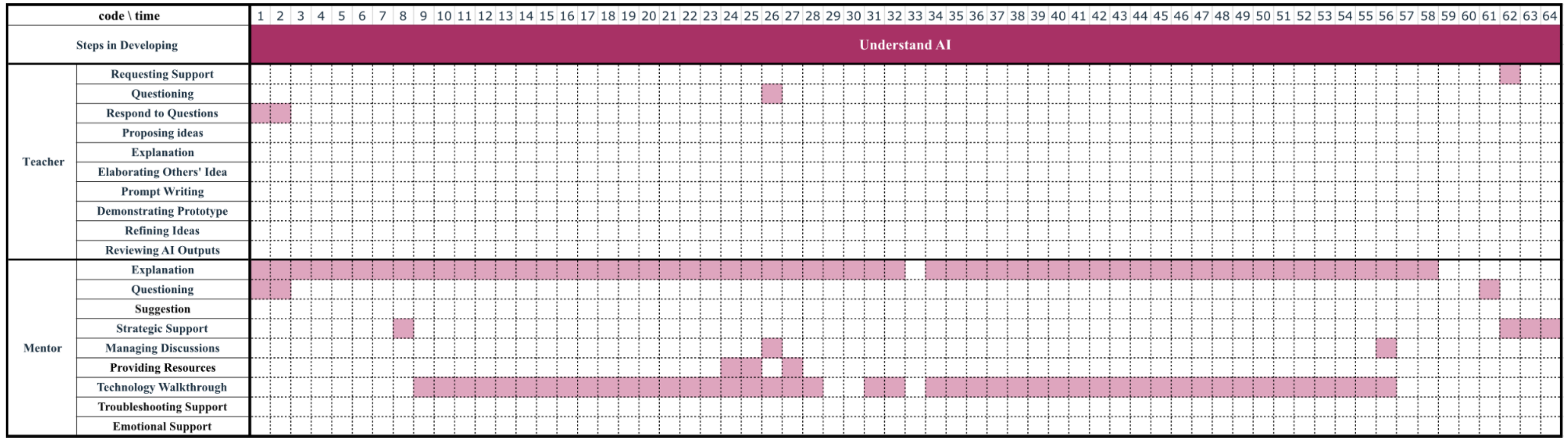

*(b) Session 3*

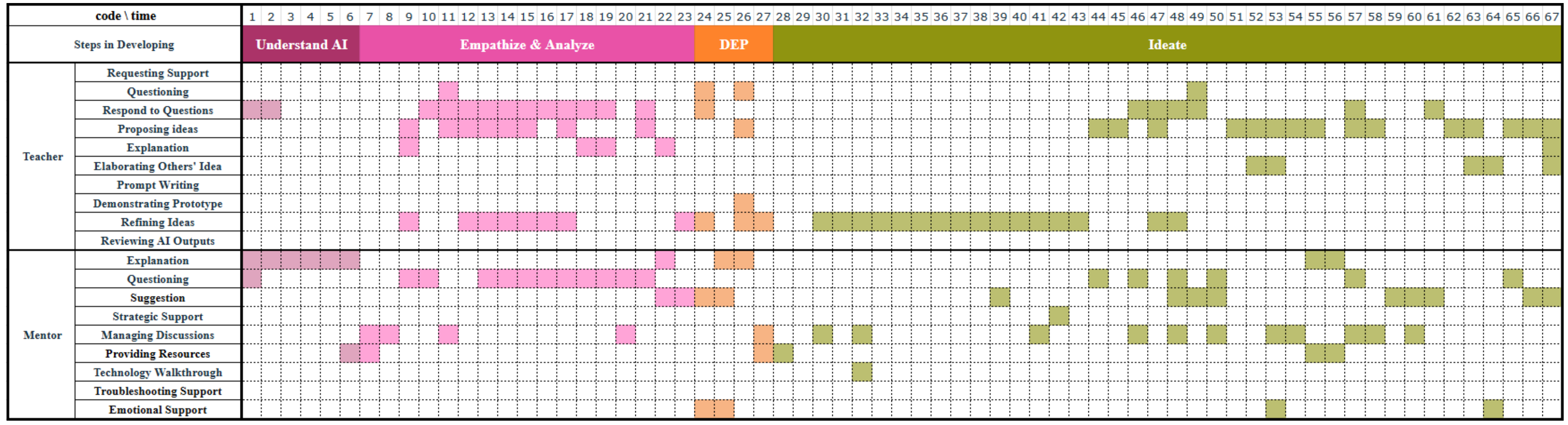

*(c) Session 4*

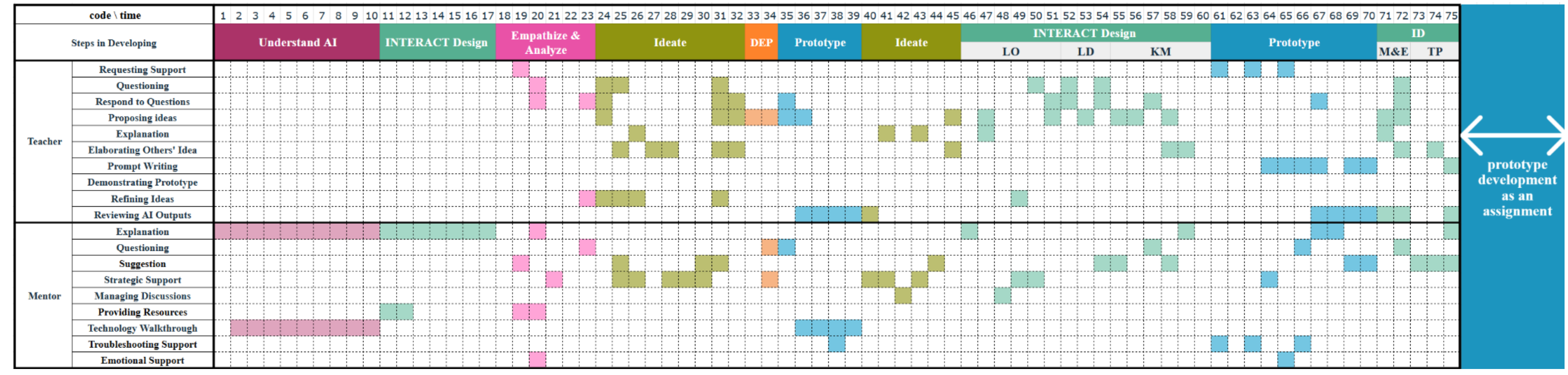

(d) Session 5, 6, 7

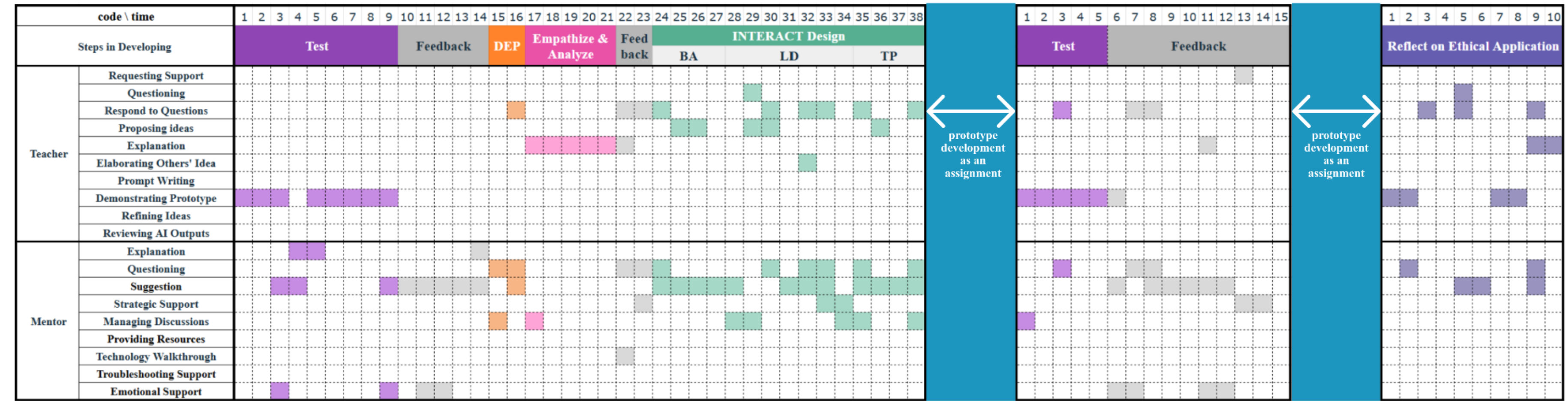

***Abbreviations used in the diagram are as follows.***

***DEP**: Define Educational Problem, **ID**: INTERACT Design, **LO**: Learning Objectives,*
***BA**: Behavioral Activities, **M&E**: Motivation & Engagement, **KM**: Knowledge Mapping,*
***LD**: Learner Differentiation, **TP**: Thinking Process*

To better understand teachers' perspectives on their technology creation process, we conducted a post-interview in which participants were asked to reflect on the visualized framework (Fig. 1) in relation to their design experiences. Overall, participants indicated that the framework was generally representative of their design process. However, they noted that some terminology, particularly concepts derived from the INTERACT framework, was not intuitive or easy for teachers to understand. To address this, we systematically reviewed each component of the INTERACT framework with participants and gathered their feedback and suggestions for revision (Table 2).

**Table 2**

*Revision of the INTERACT framework*

| Original Term | Definition | Teacher-Focused Revision | Rationale |
|---|---|---|---|
| **Behavioral Activities** | Observable physical actions that learners make to interact with the learning system. | *Keep as is* | "Learning activities" suggested by Teacher B, but we decided to keep the original term because "*behavioral activities" include all the observable activities, while learning activities could only include learning, and it makes it not differentiated from cognitive activities* (T#C) |
| **Motivation & Emotion** | Learners drive to learn, and emotional conditions of the learner that arise from the given situation. | ***Motivation & Engagement*** | *Engagement signals what we, as teachers, can design for, and it's easier to measure and observe than Emotion* (T#C) |
| **Mental Model** | Existing knowledge structures that the learner brings to the learning activity and learning gains as a result of the learning activity. | ***Knowledge mapping*** | *Mental model is a bit theoretical and Knowledge mapping is a teacher-friendly language* (T#B) |

| **Learner Variables** | Different sets of learners' cognitive and metacognitive characteristics. | ***Learner Differentiation*** | *Learner differentiation is a more teacher-friendly phrasing that evokes Universal Design for Learning* (T#B, C) |
|---|---|---|---|
| **Cognitive Activities** | Mental procedures of selecting, organizing, and integrating new information into a coherent knowledge structure. | ***Thinking Processes*** | *Cognitive is accurate but less intuitive, as teachers plan for thinking routines and strategies* (T#A, B) |
| **Learning Environment** | The situation in which the learners are spectulated to be. | *Removed* | Discussions around the Learning Environment were not observed |
| **Learning Objective** | Measurable statements of the knowledge, skills, and behaviours learners are expected to demonstrate by the end of the learning activities | *Added* | Teachers actively discussed the learning objectives and standards prior to designing learning content. |

***T#A, B, C indicates participating teachers***

Figure 4 illustrates the process model derived empirically from the workshop through CORDTRA and interview analysis. In addition to the components derived from the Design Thinking and INTERACT framework, we added *Understand AI* and *Reflect on Ethical Application* surrounding the process as they emerged in the CORDTRA analysis. The orange arrows represent the iterative pathways observed in the CORDTRA diagrams. The components in the INTERACT design phase were adapted based on the application result, with *Learning Environment* removed and *Learning Objectives* newly added.

**Figure 4**

*Process Model from the Empirical Workshop*

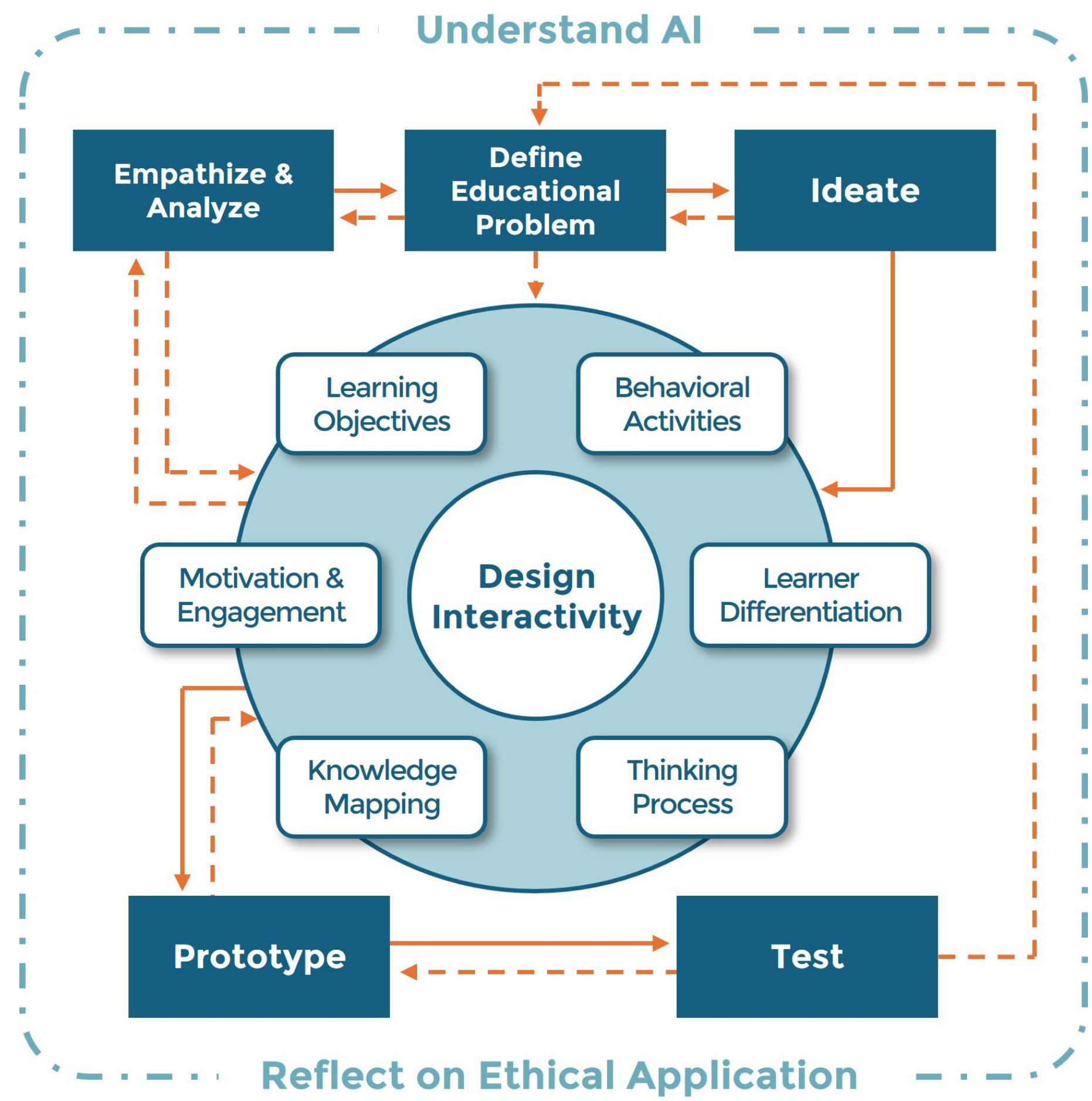

### 4.2. Activities demonstrated by teachers and mentors

While "Steps in Developing" shows the macro-level design and development process, we further analyzed the micro-level activities demonstrated by teachers and mentors (Table 3). Overall, teacher interactions accounted for 39.8%, while mentor interactions accounted for 60.2%. Within teacher interactions, *Respond to Questions* (8.9%) and *Proposing Ideas* (8.1%) were the most frequent activities, followed by *Refining Ideas (6%)*, *Explanation* (3.4%), *Demonstrating*

*Prototype* (3.3%), and *Questioning* (2.7%). In mentor interactions, *Explanation* (17%) was the most dominant activity, followed by *Technology Walkthrough* (10.7%) and *Suggestion* (10.1%).

**Table 3**
*Descriptive Analysis of Interaction Codes*

| Categories | Explanation | Frequency | % of total |
|---|---|---|---|
| **Teacher Interactions** | Respond to Questions | 49 | 8.9% |
| | Proposing ideas | 45 | 8.1% |
| | Refining Ideas | 33 | 6.0% |
| | Explanation | 19 | 3.4% |
| | Demonstrating Prototype | 18 | 3.3% |
| | Elaborating Others' Idea | 16 | 2.9% |
| | Questioning | 15 | 2.7% |
| | Reviewing AI Outputs | 12 | 2.2% |
| | Prompt Writing | 7 | 1.3% |
| | Requesting Support | 6 | 1.1% |
| **Teacher Total** | | 220 | 39.8% |
| **Mentor Interactions** | Explanation | 94 | 17% |
| | Technology Walkthrough | 59 | 10.7% |
| | Suggestion | 56 | 10.1% |
| | Questioning | 42 | 7.6% |
| | Managing Discussions | 28 | 5.1% |
| | Strategic Support | 23 | 4.2% |
| | Emotional Support | 14 | 2.5% |
| | Providing Resources | 13 | 2.4% |
| | Troubleshooting Support | 4 | 0.7% |
| **Mentor Total** | | 333 | 60.2% |
| **Total** | | 553 | 100% |

### 4.3. Evolvement of Teachers' AI Literacy

Appendix B presents the results of a thematic analysis of teachers' AI. Across four domains of AI literacy, including *Knowing and Understanding AI, Using and Applying AI, Evaluating and Creating AI,* and *AI Ethics* (Ng et al., 2021), teachers moved from initial uncertainty and instrumental views of AI toward more sophisticated, pedagogically grounded, and ethically informed perspectives.

#### *4.3.1. Know and Understand AI: From Limited Procedural Understanding to Operational Competence in AI Tools*

Before the workshop, teachers had limited procedural understanding of AI tools. While they were aware of AI applications, they lacked practical knowledge for instructional use. After the workshop, teachers reported improved functional and operational understanding, including AI's capabilities, affordances, and limitations. They developed a clearer sense of what AI can and cannot do (*"I thought AI could do everything; now I see it generates ideas but not always accurate answers"*, T#A), moved from inflated assumptions to realistic functional awareness (*"I used to think AI was like a magic box; now I see it responds to prompts and data"*, T#B), gained tool-specific understanding (*"Different AI tools have different strengths, so you need to choose the right one for the task"*, T#C), articulated mechanistic reasoning (*"It's not thinking like a human; it predicts the next word based on patterns"*, T#C), and adopted precise technological vocabulary (*"I now use terms like prompts, training data, and bias when discussing AI"*, T#B).

#### *4.3.2. Know and Understand AI: From Instrumental Views of AI to Epistemic Reframing of AI in Education*

Before the workshop, teachers primarily saw AI as a technical tool for automating tasks or generating content. After the workshop, they developed a deeper epistemic understanding of AI's role in shaping teaching and learning, shifting from abstract or polarized views to more contextualized, profession-specific interpretations (*"I didn't really know what it meant for my teaching,"* T#A). Teachers also began to view AI as instructional support rather than a threat: *I used to worry that AI might replace parts of what teachers do, but now I see it as something that* *can assist in designing learning activities"* (T#C). AI was reframed from an external tool to a collaborative partner in the learning process.

#### *4.3.3. Use and Apply AI: From Ad Hoc Experimentation to Pedagogically Meaningful AI Integration*

Teachers' approaches to integrating AI evolved from ad hoc experimentation to purpose-driven instructional design. Initially, AI use lacked pedagogical alignment ("*I tried AI once or twice, but I wasn't sure how it fit into my lesson design,"* T#C). After the workshop, teachers reported clearer alignment with learning objectives (*"Now I start by thinking about the learning objective first, then consider how AI might support it,"* T#C), integration with existing instructional technologies (*"I'm now thinking about how AI can work alongside the platforms we already use,"* T#C), and focus on instructional value rather than novelty (*"It's not about using AI because it's new; it's about whether it helps students learn,"* T#A). These shifts reflect a move from surface-level adoption to intentional, goal-oriented AI integration.

#### *4.3.4. Use and Apply AI: From AI as a Substitute Tool to Human-AI Collaborative Instruction*

While teachers viewed AI as a potential substitute for teachers before the workshop, they began to reconceptualize AI as a collaborative partner, augmenting rather than replacing human expertise.

> "I realized AI isn't replacing what I do as a teacher. It's more like a partner I can work with. I can ask it for ideas, refine the responses, and then adapt them to design activities that fit my students." (T#C)

#### *4.3.5. Use and Apply AI: From Instructional Uncertainty to Confidence in AI-Supported Teaching*

Teachers also reported increased confidence in implementing AI-supported instruction. Initially, participants expressed hesitation and uncertainty about using AI in the classroom, as noted by T#B in Appendix A. Following the workshop, however, teachers described reduced hesitation in experimenting with AI tools (*"After the workshop, I feel more comfortable trying AI tools in my lessons and seeing how they work with my students."*). Teachers also reported greater confidence in selecting AI tools for specific instructional tasks (*"Now I feel more confident choosing when and how to use AI depending on what I want students to accomplish*," T#B).

#### ***4.3.6. Evaluate and Create AI: From Limited Awareness to Critical Understanding of AI's Sociotechnical Implications***

Another important shift involved teachers' awareness of the broader sociotechnical implications of AI. Before the workshop, participants rarely considered how AI might reflect societal biases or cultural assumptions. After the workshop, teachers demonstrated greater critical awareness of how AI integrates into daily life, reshapes human performance, and influences sociocultural dynamics in schools.

> "*I realize now that AI isn't neutral; it affects how people perform tasks, make decisions, and even perceive knowledge. It actually influences the social and cultural context of learning, such as equity, representation, and the hidden assumptions built into the technologies we use*." (T#C)

#### ***4.3.7. Evaluate and Create AI: From Passive Consumption to Teacher Agency as Designers of AI-Supported Learning***

Teachers' roles in relation to AI shifted from passive users to active creators. Initially, they primarily engaged with pre-existing tools, but after the workshop, they began designing AI-supported learning experiences, taking ownership of integration decisions, and adapting or creating tools to fit classroom needs: *I used to just follow whatever AI tool was suggested, but now I feel confident modifying and designing AI-driven learning activities that truly fit my students' need* (T#A).

#### *4.3.8. AI Ethics: From Ambiguous Awareness to Nuanced Understanding of AI Capabilities and Limitations*

Teachers' ethical awareness of AI became more nuanced. Before the workshop, they expressed general concerns about AI reliability but had limited understanding of its limitations. After the workshop, they demonstrated greater awareness of responsible AI use, including potential biases, hallucinations, transparency issues, risks of over-reliance, and the intended and unintended consequences of underuse, misuse, or overuse of AI.

> *I used to focus on whether AI worked, but now I also consider its limitations, transparency, and potential unintended effects in the classroom. Understanding AI's ethical boundaries helps me make informed decisions about which tools to use and how to integrate them safely and effectively.* (T#C)

#### *4.3.9. AI Ethics: From Compliance-Based Concerns to Pedagogically Purposeful AI Use*

Teachers' ethical considerations shifted from rule-based concerns about misuse to the proactive integration of ethical practices into pedagogy. They now intentionally align AI use with educational values, critically reflecting on fairness, bias, and instructional appropriateness, as illustrated by T#B: *"Now I consider bias and instructional appropriateness before integrating any AI tool, not just whether it works"*.

#### *4.3.10. AI Ethics: From Unclear Boundaries to Intentional Delineation of Human-AI Roles*

Finally, teachers developed clearer strategies for assigning roles to humans and AI in learning activities. Initially uncertain about balancing AI assistance with student responsibilities, they later intentionally designed tasks where AI supported specific processes while students and teachers retained responsibility for critical thinking and interpretation, emphasizing the preservation of human connections and uniquely human capacities such as empathy and moral guidance.

*"I used to let AI handle tasks automatically, but now I decide which parts AI handles and which require human judgment, ensuring empathy and guidance remain the teacher's role and AI supports learning without replacing human interaction."* (T#C)

# V. Discussion

## 5.1. GAIDE : A Guiding Framework for AI-Integrated Design for Educators

After the empirical application of the initial framework, Figure 5 presents the revised final framework, *GAIDE: A Guiding Framework for AI-Integrated Design for Educators*.

**Figure 5**

*GAIDE framework (A Guiding framework for AI-integrated Design for Educators)*

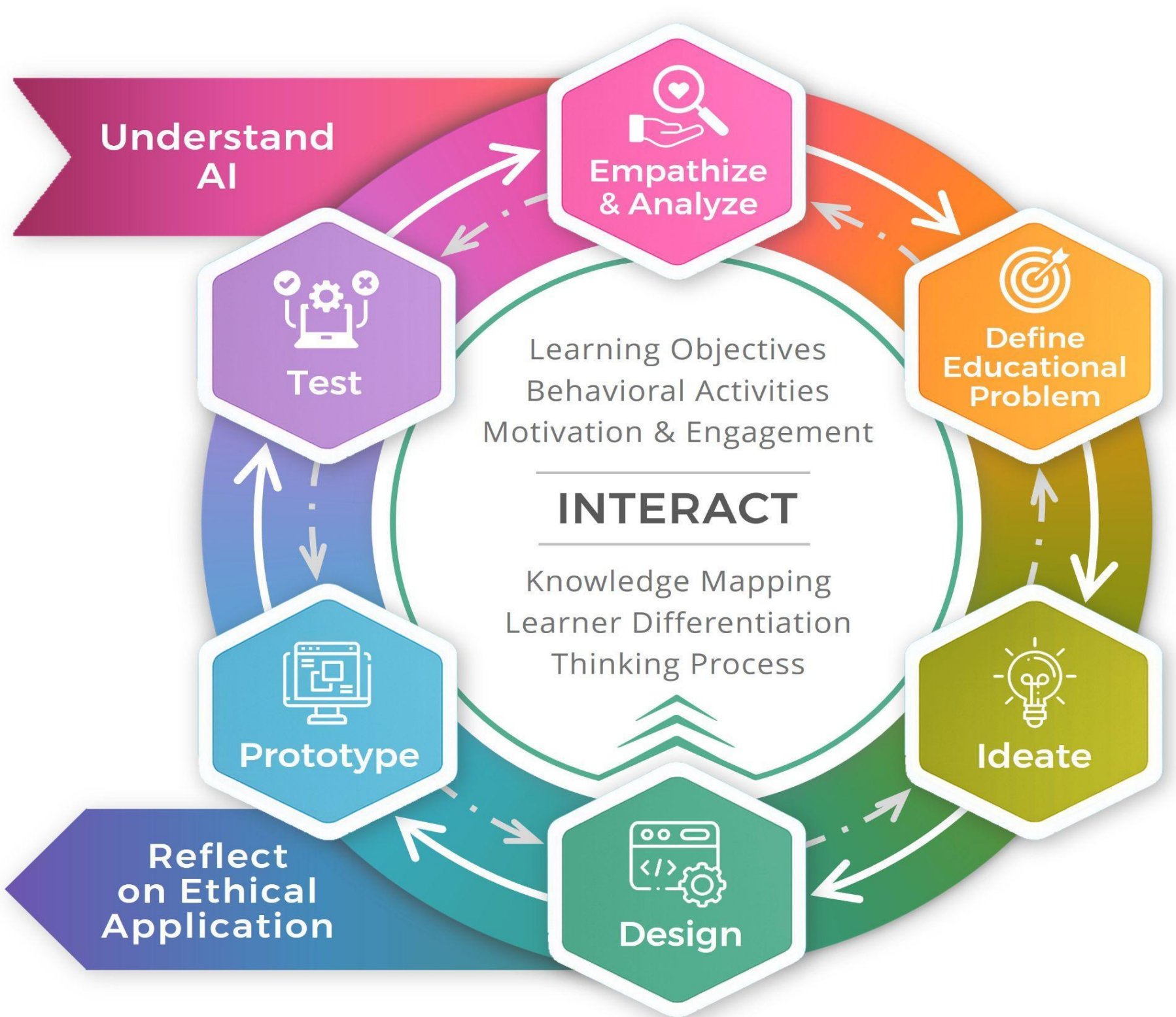

The framework begins with establishing a foundational understanding of AI prior to engaging in the design and development process. This aligns with existing conceptualizations of AI literacy (e.g., Ng et al., 2021) as well as prior work on the design and implementation of teacher professional development (Pei et al., 2025). Following this, the framework incorporates stages inspired by Design Thinking (Razzouk & Shute, 2012). Reflecting our analysis, our framework visualizes non-linear but iterative cyclic process of interpretation and negotiation during the teachers' collaborative design (Clark & Hollingsworth, 2002). Within the *Design* phase, which integrates components of the INTERACT framework, several adaptations were made based on our empirical implementation: *Learning Environment* was removed, and *Learning Objectives* was newly introduced. Although teachers were explicitly guided through the INTERACT design components, results indicated limited engagement with discussions of the learning environment. One possible explanation is that considerations of learning environment were implicitly addressed during the *Empathize* phase of Design Thinking, where teachers naturally reflected on learners' environmental contexts (Razzouk & Shute, 2012). In contrast, *Learning Objectives* was added to better align with teachers' design practices. Although *Learning Objectives* is not explicitly included in the original INTERACT framework, defining learning objectives is a critical step prior to developing instructional materials, as emphasized in many instructional design models (e.g., Dick et al., 2015). This addition extends the framework beyond a general Design Thinking approach by grounding it more explicitly in instructional design. Following the *Design* phase, teachers engaged in *Prototyping* and *Testing* their solutions with mentors, receiving feedback, and iteratively refining their designs. Finally, teachers reflected on the ethical implications and responsible application of their developed AI-powered tools, thereby closing the loop with AI literacy development (Ng et al., 2021).

Our framework advances teachers' roles from *co-designers* to *primary designers.* While prior co-design and participatory design studies (e.g., Kali et al., 2015) have involved teachers in the development of educational technologies, these collaborative models have often positioned researchers or technical developers as the primary drivers of the design process. In these models, teachers had limited contributions, such as domain expertise or partial participation, but rarely engaged in the entire process and played main roles in the technological design and development (Lin & Van Brummelen, 2021; Nicholson, 2022). Our framework attempts to overcome this challenge by positioning teachers as the primary designers, supported by AI-enabled vibe coding. This shift grants teachers greater agency and enables them to more directly shape technologies to align with their instructional goals.

### 5.2. AI-Teacher-Mentor Collaboration Synergy in AI Learning Technology Creation

Through fine-grained interaction analysis of teachers and mentors using the CORDTRA diagram, we conceptualize the interactions among teachers, mentors, and AI as a dynamically coordinated system of distributed expertise, grounded in distributed cognition theory (Putnam & Borko, 2000). In this system, evidence-based knowledge from mentors, practice-based knowledge from teachers, and generative knowledge from AI are synergistically integrated and continuously reconfigured through interaction within the situated context of authentic classroom practice (Voogt et al., 2015).

First, teachers exercise agency by actively engaging in design and implementation, demonstrating their potential to initiate educational change (Voogt et al., 2015). We observed active dialogic interactions among teachers throughout the design and development process. *Responding to Questions* (8.9%) and *Proposing Ideas* (8.1%) were the most frequent, indicating strong engagement in collaborative idea exchange. *Refining Ideas* (6%), particularly prominent during *Empathize & Analyze* and *Ideate*, reflects deep cognitive and metacognitive engagement. *Explanation* (3.4%), *Elaborating Others' Ideas* (2.9%), and *Questioning* (2.7%) further illustrate sustained collaborative interaction across phases. Such human-human dialogic interactions have been widely recognized as central to effective co-design of instructional materials and teacher professional development, as collaborative design itself constitutes an authentic learning context for teachers (McKenney et al., 2016).

Beyond human collaboration, our findings suggest AI can be an additional and integral actor in the design process. Teachers' direct interactions with AI, including *Demonstrating Prototypes* (3.3%), *Reviewing AI Outputs* (2.2%), and *Prompt Writing* (1.3%), were observed, suggesting that AI contributes generative knowledge that supports distributed cognition (Hollan et al., 2000) by producing code, generating working prototypes, and offering novel ideas. Within the vibe coding environment, teachers leveraged AI to articulate, expand, and refine their ideas, externalizing tacit knowledge and engaging in iterative cycles of generation, testing, and revision, highlighting learning as occurring through cognitive mediation with AI. The teacher interview further revealed this perspective, as teacher C said they perceive AI as both a cognitive offloader (Choi et al., 2026) and a collaborative partner (Sun & Huang, 2025). Building on this perspective, our findings suggest that AI should be understood not merely as a tool, but as a mediational artifact that actively shapes teachers' thinking processes (Voogt et al., 2015).

Finally, analysis of mentors' activities provides insight into effective mentoring practices in AI-supported technology design. Mentors' support activities can be categorized into three domains by Gagné (1985)'s theory: cognitive (knowledge), psychomotor (skills), and affective (attitudes). Cognitive support included *Explanation*, *Suggestion*, *Questioning*, *Managing Discussions*, *Strategic Support*, and *Providing Resources*. Notably, *Questioning* (7.6%) and *Managing Discussions* (5.1%) were observed across all phases, indicating sustained facilitation of teachers' thinking and coordination of collaborative processes. *Strategic Support* (4.2%), particularly during the *Ideate* phase, helped teachers refine and operationalize ideas into feasible design directions. In addition, mentors supported skill development through activities such as *Technology Walkthroughs* and *Troubleshooting* to support technical skills with AI. Underscoring the importance of comprehensive mentoring, the mentoring patterns reflect principles of cognitive apprenticeship (Collins, 2013), including modeling, coaching, and scaffolding with gradual fading of support. In the workshop, mentors made expert thinking visible, guided teachers' reasoning through questioning and feedback, and progressively reduced support to enable teachers to assume greater ownership and agency in the design process.

### 5.3. Learning-by-Creating AI: A Teacher Professional Development Opportunity

From the interview, we identified evolving themes across key components of AI literacy. These shifts suggest that engaging in the AI creation process can serve as a meaningful form of professional development for teachers' AI learning (Hsu et al., 2017; Kalantzis & Cope, 2005; Papert, 1980). The findings can further be interpreted through the lens of the TPACK framework (Mishra et al., 2023). The workshop activities strengthened teachers' Technological Knowledge (TK) by providing hands-on exposure to AI tools and Technological Pedagogical Knowledge (TPK) by having them explore how AI could be integrated into their instructional practices. Importantly, the process of designing and creating AI artifacts shifted teachers' perspectives from viewing AI as a tool for isolated use toward understanding it as a resource for strategic pedagogical integration in authentic classroom contexts, thereby supporting the development of Technological Pedagogical Content Knowledge (TPACK). This progression is also reflected in the final stage of the framework, *Reflect on Ethical Application,* where teachers engaged in discussions about the ethical implications of implementing AI tools in classroom settings, reinforcing the role of ethical reasoning as a core component of AI literacy (Ng et al., 2021).

Additionally, the findings highlight teachers' emphasis on human-centered competencies, including the ability to empathetically identify meaningful educational problems, foster authentic teacher-student interactions, promote students' critical thinking and creativity in AI-supported contexts, and exercise moral and ethical judgment in the use of AI. These competencies should no longer be regarded as "soft skills" but critical competencies in the age of AI (OECD, 2019). As AI expands what is technically possible, teachers have consistently emphasized the importance of leveraging AI to augment human capacities rather than replacing them.

## VI. Conclusion

While AI-assisted vibe coding has democratized software development, human insight for identifying meaningful problems with contextual expertise is becoming even more important (OECD, 2019). Teachers, in particular, are well-positioned to recognize these needs within authentic learning environments; This capacity must be intentionally supported and cultivated. To address this need, this paper proposes GAIDE (A Guiding framework for AI-integrated Design for Educators). The integration of Design Thinking, especially its emphasis on empathy, serves as a critical mechanism for foregrounding this human-centered perspective (Razzouk & Shute, 2012), while the incorporation of INTERACT design components ensures that learner agency remains central to the design of pedagogically meaningful learning experiences (Domagk et al., 2010). We examined the usability and practicability of our framework in an authentic learning workshop setting and reflected on the revisions to derive the final GAIDE framework. While more user cases are needed, our framework serves as a practical guide for teachers to leverage their contextual knowledge and empathetic understanding to identify problems, develop AI-powered solutions, and engage in purposeful, human-centered innovation in the age of AI.

## References

Alenezi, M., & Akour, M. (2025). Ai-driven innovations in software engineering: a review of current practices and future directions. *Applied Sciences, 15*(3), 1344. https://doi.org/10.3390/app15031344

Braun, V., & Clarke, V. (2006). Using thematic analysis in psychology. *Qualitative research in psychology, 3*(2), 77-101. https://doi.org/10.1191/1478088706qp063oa

Celik, I., Dindar, M., Muukkonen, H., & Järvelä, S. (2022). The promises and challenges of artificial intelligence for teachers: A systematic review of research. *TechTrends, 66*(4), 616-630. https://doi.org/10.1007/s11528-022-00715-y

Choi, S., Jeong, S., Park, S., Kim, Y., & Han, I. (2026). Analyzing teacher–AI interaction patterns across teacher experience and AI proficiency in student-centered lesson design. *Teaching and Teacher Education, 169*, 105266. https://doi.org/10.1016/j.tate.2025.105266

Chow, M., & Ng, O. (2025). From technology adopters to creators: Leveraging AI-assisted vibe coding to transform clinical teaching and learning. *Medical teacher, 47*(12), 1927-1929. https://doi.org/10.1080/0142159X.2025.2488353

Collins, A. (2013). Cognitive apprenticeship and instructional technology. In *Educational values and cognitive instruction* (pp. 121-138). Routledge.

Dick, W., Carey, L., & Carey, J. O. (2015). The Systematic Design of Instruction (8th ed.). Upper Saddle River, NJ: Pearson.

Ding, A. C. E., Shi, L., Yang, H., & Choi, I. (2024). Enhancing teacher AI literacy and integration through different types of cases in teacher professional development. *Computers and Education open*, *6*, 100178. https://doi.org/10.1016/j.caeo.2024.100178

Domagk, S., Schwartz, R. N., & Plass, J. L. (2010). Interactivity in multimedia learning: An integrated model. *Computers in Human Behavior*, *26*(5), 1024-1033. https://doi.org/10.1016/j.chb.2010.03.003

Fereday, J., & Muir-Cochrane, E. (2006). Demonstrating rigor using thematic analysis: A hybrid approach of inductive and deductive coding and theme development. *International journal of qualitative methods, 5*(1), 80-92. https://doi.org/10.1177/160940690600500107

Gagné, R. M. (1985). *The conditions of learning and theory of instruction* (4th ed.). New York, NY: Holt, Rinehart & Winston.

Hmelo-Silver, C. E., & Barrows, H. S. (2006). Goals and strategies of a problem-based learning facilitator. *Interdisciplinary journal of problem-based learning*, *1*(1), 4. https://doi.org/10.7771/1541-5015.1004

Hmelo-Silver, C. E., Liu, L. E. I., & Jordan, R. (2009). Visual representation of a multidimensional coding scheme for understanding technology-mediated learning about complex natural systems. *Research and Practice in Technology Enhanced Learning, 4*(3), 253-280. https://doi.org/10.1142/S1793206809000714

Hollan, J., Hutchins, E., & Kirsh, D. (2000). Distributed cognition: toward a new foundation for human-computer interaction research. *ACM Transactions on Computer-Human Interaction (TOCHI)*, *7*(2), 174-196. https://doi.org/10.1145/353485.353487

Hsu, Y. C., Baldwin, S., & Ching, Y. H. (2017). Learning through making and maker education. *TechTrends*, *61*(6), 589-594. https://doi.org/10.1007/s11528-017-0172-6

Kalantzis, M., & Cope, B. (2005). *Learning by design*. Common Ground.

Kali, Y., McKenney, S., & Sagy, O. (2015). Teachers as designers of technology enhanced learning. *Instructional science*, *43*(2), 173-179. https://doi.org/10.1007/s11251-014-9343-4

Karpathy, A. (2025). Vibe coding. Post on X (formerly Twitter), [accessed on March 15th, 2026]. retrieved from https://x.com/karpathy/status/1886192184808149383

Koehler, M. J., & Mishra, P. (2005). What happens when teachers design educational technology? The development of technological pedagogical content knowledge. *Journal of educational computing research*, *32*(2), 131-152. https://doi.org/10.2190/0EW7-01WB-BKHL-QDYV

Kusper, G., & Szabó, C. (2025, November). Vibe Coding in Education. In 2025 International Conference on *Emerging eLearning Technologies and Applications* (ICETA) (pp. 506-511). IEEE. 10.1109/ICETA67772.2025.11280285

Lin, P., & Van Brummelen, J. (2021, May). Engaging teachers to co-design integrated AI curriculum for K-12 classrooms. In *Proceedings of the 2021 CHI conference on human factors in computing systems* (pp. 1-12). https://doi.org/10.1145/3411764.3445377

Long, D., & Magerko, B. (2020, April). What is AI literacy? Competencies and design considerations. In *Proceedings of the 2020 CHI conference on human factors in computing systems* (pp. 1-16). https://doi.org/10.1145/3313831.3376727.

Luckin, R., & Cukurova, M. (2019). Designing educational technologies in the age of AI: A learning sciences-driven approach. *British Journal of Educational Technology, 50*(6), 2824-2838. https://doi.org/10.1111/bjet.12861Digital

Martin, F., Wang, C., & Sadaf, A. (2020). Facilitation matters: Instructor perception of helpfulness of facilitation strategies in online courses. *Online Learning, 24*(1), 28-49. https://doi.org/10.24059/olj.v24i1.1980

McKenney, S., Boschman, F., Pieters, J., & Voogt, J. (2016). Collaborative design of technology-enhanced learning: What can we learn from teacher talk?. *TechTrends*, *60*(4), 385-391. https://doi.org/10.1007/s11528-016-0078-8

Minervee. (2026, February 25). *Claude Hackathon Winners: 5 Non-Coder Apps in One Week*. Medium; Coding Nexus. https://medium.com/coding-nexus/claude-hackathon-winners-5-non-coder-apps-in-one-week-736425986774

Mishra, P., & Koehler, M. J. (2006). Technological pedagogical content knowledge: A framework for teacher knowledge. *Teachers college record, 108*(6), 1017-1054. https://doi.org/10.1111/j.1467-9620.2006.00684.x

Mishra, P., Warr, M., & Islam, R. (2023). TPACK in the age of ChatGPT and Generative AI. *Journal of Digital Learning in Teacher Education, 39*(4), 235-251. https://doi.org/10.1080/21532974.2023.2247480

Ng, D. T. K., Leung, J. K. L., Chu, S. K. W., & Qiao, M. S. (2021). Conceptualizing AI literacy: An exploratory review. *Computers and Education: Artificial Intelligence,* 2, 100041. https://doi.org/10.1016/j.caeai.2021.100041

Nicholson, R., Bartindale, T., Kharrufa, A., Kirk, D., & Walker-Gleaves, C. (2022, April). Participatory design goes to school: Co-teaching as a form of co-design for educational technology. In *Proceedings of the 2022 CHI conference on human factors in computing systems* (pp. 1-17). https://doi.org/10.1145/3491102.3517667

Organization for Economic Co-operation and Development (OECD). (2019). *OECD skills outlook 2019: Thriving in a digital world*. OECD Publishing. https://doi.org/10.1787/df80bc12-en

Papert, S. (1980). *Mindstorms: Children, computers, and powerful ideas*. Basic Books.

Pei, B., Lu, J., & Jing, X. (2025). Empowering preservice teachers' AI literacy: Current understanding, influential factors, and strategies for improvement. *Computers and Education: Artificial Intelligence*, 8, 100406. https://doi.org/10.1016/j.caeai.2025.100406

Penuel, W. R., Fishman, B. J., Yamaguchi, R., & Gallagher, L. P. (2007). What makes professional development effective? Strategies that foster curriculum implementation. *American educational research journal, 44*(4), 921-958. https://doi.org/10.3102/0002831207308221

Putnam, R. T., & Borko, H. (2000). What do new views of knowledge and thinking have to say about research on teacher learning?. *Educational researcher, 29*(1), 4-15. https://doi.org/10.3102/0013189X029001004

Razzouk, R., & Shute, V. (2012). What is design thinking and why is it important?. *Review of educational research, 82*(3), 330-348. https://doi.org/10.3102/0034654312457429

Richey, R. C., & Klein, J. D. (2013). Design and development research. In Handbook of *research on educational communications and technology* (pp. 141-150). New York, NY: Springer New York.

Roschelle, J., Penuel, W., & Shechtman, N. (2006). Co-design of innovations with teachers: Definition and dynamics. In *Proceedings of* the *2006 International Conference of the Learning Sciences (ICLS)* (pp. 606-612). https://doi.dx.org/10.22318/icls2006.606

Song, Y., Li, C., Ma, Y., Lyu, B., Zhu, W., Li, H., & Xing, W. (2025). Exploring effective tutoring strategies in asynchronous online mathematical discussions: Insights from ordered network analysis. *Journal of Science Education and Technology*, *34*(5), 1143-1163. https://doi.org/10.1007/s10956-025-10233-0

Song, Y., Wong, R., Kim, J., Moon, J., Patton, E., Phan, V., McCullum, J., & Day, J. (2026, in press). Supporting K-12 Teachers in the Presidential AI Challenge: A Case Study of a Faculty-Mentored Workshop for AI Tool Creation. In *Proceedings of The 27th International Conference on Artificial Intelligence in Education (AIED)*.

Sun, Y., & Huang, Y. (2025). Exploring the Role of Generative AI in Collaborative Lesson Planning for Pre-Service Teachers. *Journal of Computer Assisted Learning, 41*(4), e70098. https://doi.org/10.1111/jcal.70098

The White House, Office of Science and Technology Policy, & National Science and Technology Council. (2025). *Presidential AI Challenge Guidebook for Participation.* [accessed on January 15th, 2026]. Retrieved from https://www.whitehouse.gov/wp-content/uploads/2025/08/Presidential-AI-Challenge-Guidebook-for-Participation.pdf

Topping, K., Miller, D., Murray, P., & Conlin, N. (2011). Implementation integrity in peer tutoring of mathematics. *Educational Psychology*, *31*(5), 575-593. https://doi.org/10.1080/01443410.2011.585949

Touretzky, D., Gardner-McCune, C., Martin, F., & Seehorn, D. (2019, July). Envisioning AI for K-12: What should every child know about AI?. In *Proceedings of the AAAI conference on artificial intelligence* (Vol. 33, No. 01, pp. 9795-9799). https://doi.org/10.1609/aaai.v33i01.33019795

Voogt, J., Laferriere, T., Breuleux, A., Itow, R. C., Hickey, D. T., & McKenney, S. (2015). Collaborative design as a form of professional development. *Instructional science, 43*(2), 259-282. https://doi.org/10.1007/s11251-014-9340-7

Woo, D. J., Guo, K., & Yu, Y. (2025). A vibe coding learning design to enhance EFL students' talking to, through, and about AI. *arXiv preprint arXiv:2509.08854.*

Yan, J., & Goh, H. H. (2023). Exploring the cognitive processes in teacher candidates' collaborative task-based lesson planning. *Teaching and Teacher Education*, *136*, 104365. https://doi.org/10.1016/j.tate.2023.104365

**Declaration of interest statement**

The authors declare that they have no conflicts of interest.

**Appendix A**

*Coding Scheme of Teacher and Mentor Interactions*

| Category | Code (References) | Definition | Examples |
|---|---|---|---|
| **Teacher Interactions** | Requesting Support (Yan & Goh, 2023) | Requesting technical or cognitive assistance. | “All right, so what should we put in… I’m struggling here.” “Can you help me with some ideas?” |
| | Questioning (Yan & Goh, 2023) | Asking questions to seek information or clarification. | “Do we want to focus on the ELA standards or do we want to go back into science?” |
| | Respond to Questions (*D) | Providing responses with general information without introducing new ideas. | “We named her ‘Maddie’” [persona name] “Oh yeah, we can upload math standards.” |
| | Proposing ideas (*D) | Introducing new ideas as the main focus of the utterance. | “Maybe we target science.” “What if we went old school and lumped all of our standards under Bloom's hierarchy…?” |
| | Explanation (*D) | Explaining one’s own thinking or reasoning. | “The standards have multiple parts… we can pick and choose among them.” |
| | Elaborating Others' Idea (Yan & Goh, 2023) | Building on others’ ideas by adding or refining them. | “And then she could cross-reference between the different apps.” |
| | Prompt Writing (*D) | Writing prompts to interact with AI. | The teacher types a prompt in the AI studio to generate or modify the output. |
| | Demonstrating Prototype (*D) | Presenting or explaining a created artifact or prototype. | The teacher presents the prototype on the screen and explains how its features work. |
| | Refining Ideas (*D) | Organizing and refining ideas from the discussion using digital tools or shared platforms. | The teacher organizes and records ideas from the discussion in a shared document. |
| | Reviewing AI Outputs (*D) | Examining and evaluating AI-generated responses. | The teacher reviews the AI output and determines whether it is appropriate or needs revision. |
| **Mentor Interactions** | Explanation (Pol et al., 2010; Song et al., 2025) | Explaining concepts or procedures to support the teacher’s understanding. | “So, we can think about what machines and human intelligence do.”, “You can see there are multiple buttons here. But the major things that you are going to interact here is this chat interface.” |
| | Questioning (Hmelo-Silver & Barrows, 2006) | Asking questions to prompt thinking, clarification, or reflection. | “What is the purpose of this app?” |
| | Suggestion (Song et al., 2025) | Offering ideas or recommendations relevant to the teacher’s context, even in the absence of a problem. | “Why don’t we incorporate Bloom’s taxonomy into the design specifications?” |
| | Strategic Support (Hmelo-Silver & Barrows, 2006) | Providing guidance on strategies or approaches to accomplish the task effectively. | “One approach is to narrow down the scope to make it more feasible.” |

| | | |
|---|---|---|
| Managing Discussions (Song et al., 2025) | Organizing the flow of discussion or shifting topics. | "OK sounds good. Any other questions? One teacher, do you have any comments? Now, let's move to another teacher's project." |
| Providing Resources (Martin et al., 2020) | Providing the teacher with reference materials, links, or templates. | The pastes a resource link into the chat. |
| Technology Walkthrough (*D) | Demonstrating how to use technology to support the teacher's understanding. | The mentor opens an AI tool on screen and demonstrates how to use it. |
| Troubleshooting Support (*D) | Providing support to resolve problems encountered during task performance. | The mentor instructs the teacher to input the error into an AI tool to debug it. |
| Emotional Support (Song et al., 2025) | Expressing emotional support to the teacher. | "Yeah don't worry once you have a prompt, you can definitely make it your prototype within 10 minutes." |

***D indicates the code inductively drawn from the data***

## Appendix B

*Thematic Comparison of Teachers' AI Literacy Before and After AI Technology Creation*

| **Domain of AI literacy** | **Themes** | |
|---|---|---|
| | **Before workshop** | **After workshop** |
| Know and understand AI | 1. Limited procedural understanding of AI tools<br>"*I had heard about AI tools like ChatGPT, but honestly I didn't really know how they worked or how to use them effectively in my class*." (T#A) | 1. Operational competence in using and configuring AI tools and knowledge<br>*"I feel like I actually understand how to interact with AI tools, how to craft prompts and adjust outputs so they support my lesson goals."* (T#A) |
| | 2. Instrumental view of AI as a technical tool<br>*"I mostly thought of AI as just another digital tool that could help generate worksheets or quick answers."* (T#C) | 2. Epistemic reframing of AI as a knowledge-shaping educational partner<br>*"The workshop helped me realize AI can shape how students think and construct knowledge, not just produce content."* (T#C) |
| Use and apply AI | 3. Unstructured AI integration in the classroom instruction<br>*"I tried AI once or twice, but it was more like experimenting. I wasn't really sure how it fit into my lesson design."* (T#B) | 3. Pedagogically aligned and meaningful AI integration<br>*"Now I intentionally design activities where AI supports specific learning* |
| | | *objectives, rather than just adding it as a novelty."* (T#B) |
| | 4. AI as substitute or stand-alone tool<br>*"At first I thought AI might replace some teacher tasks like grading or explaining content."* (T#C) | 4. Human-AI Collaborative Capacity in instructional practice<br>*"I now see AI more as a collaborator; it can help generate ideas, but the teacher still guides the learning process."* (T#C) |
| | 5. Instructional uncertainty with AI integration<br>*"I feel hesitant to use AI in class because I am not sure if I could manage it or if it would even help students."* (T#B) | Instructional confidence in AI-supported teaching<br>*"After the workshop, I feel much more confident experimenting with AI in my lessons."* (T#B) |
| Evaluate and create AI | 6. Limited awareness of AI's sociotechnical implications<br>*"I didn't really think about the broader impacts of AI. I mostly saw it as a convenient tool."* (T#C) | Critical awareness of AI's sociotechnical and technocultural dimensions<br>*"Now I'm more aware that AI systems reflect societal biases and cultural assumptions."* (T#C) |
| | 7. Consumer-oriented engagement with AI tools<br>*"Before, I mainly used AI tools that were already built. I didn't think about creating or adapting them for my own teaching."* (T#B) | Agentic role as a co-designer of AI-supported learning experience<br>*"The workshop helped me see that teachers can actually design AI-supported learning activities tailored to our classrooms."* (T#B) |
| AI ethics | 8. Ambiguous understanding of AI capabilities and risks<br>*"I knew AI might have some issues, like giving wrong answers, but I didn't really know the details."* (T#A) | Nuanced ethical awareness of AI capabilities, biases, and limitations<br>*"Now I pay closer attention to issues like bias, hallucination, and transparency when students use AI."* (T#A) |
| | 9. Compliance-based ethical consideration<br>*"My main concern is just making sure students are not cheating with AI."* (T#C) | Pedagogically purposeful ethical use of AI in learning design<br>*"Instead of banning AI, I'm thinking about how to guide students to use it responsibly as part of the learning process."* (T#C) |
| | 10. Unclear boundaries between human and AI roles<br>*"I wasn't sure which tasks should be done by instructors and which ones AI could support."* (T#A) | Intentional delineation of Human-AI roles in educational practice<br>*"Now I clearly design tasks where students and instructors do the critical thinking while AI supports brainstorming or feedback."* (T#A) |